\documentstyle[aps,twocolumn,prl,epsfig]{revtex}

\newcommand{\be}{\begin{equation}}
\newcommand{\ee}{\end{equation}}
\newcommand{\bea}{\begin{eqnarray}}
\newcommand{\eea}{\end{eqnarray}}

\begin{document}

\tighten\draft

\twocolumn[\hsize\textwidth\columnwidth\hsize\csname@twocolumnfalse\endcsname

\title{A Bayesian Approach to Inverse Quantum Statistics}
\author{J. C.  Lemm, J. Uhlig, and A. Weiguny} 
\address{
Institut f\"ur Theoretische Physik I,
Universit\"at M\"unster, 48149 M\"unster, Germany
}
\date{\today}
\maketitle

\begin{abstract}
A nonparametric Bayesian approach is developed
to determine quantum potentials
from empirical data
for quantum systems at finite temperature.
The approach combines
the likelihood model of quantum mechanics
with a priori information over potentials
implemented in form of stochastic processes.
Its specific advantages
are the possibilities to deal with heterogeneous data
and to express a priori information explicitly, i.e.,
directly in terms of the potential of interest.
A numerical solution in maximum a posteriori approximation
was feasible for one--dimensional problems.
Using correct 
a priori information turned out to be essential.
\end{abstract}
] 

\narrowtext
The last years have seen a rapidly growing interest 
in learning from empirical data.
Increasing computational resources enabled successful applications of 
empirical learning algorithms in many different areas including, for example,
time series prediction,
image reconstruction,
speech recognition,
and many more regression, classification,
and density estimation problems.
Empirical learning, i.e.,
the problem of finding underlying general laws from observations,
represents a typical inverse problem
and is usually ill--posed in the sense of Hadamard
\cite{iqs:Tikhonov-Arsenin:1977,iqs:Vapnik:1982,iqs:Kirsch:1996}.
It is well known that a successful solution of such problems
requires additional a priori information.
In the setting of empirical learning
it is a priori information which
controls the generalization ability of a learning system
by providing the link between available empirical ``training'' data 
and unknown outcome in future ``test'' situations.

The empirical learning problem we study in this Paper
is the reconstruction of potentials  
from measuring quantum systems at finite temperature, i.e.,
the problem of inverse quantum statistics.
Two classical research fields dealing
with the determination of potentials
are inverse scattering theory
\cite{iqs:Newton:1989} 
and inverse spectral theory 
\cite{iqs:Chadan-Colton-Paivarinta-Rundell:1997,iqs:Zakhariev-Chabanov:1997}.
They characterize the kind of data which
are necessary, in addition to a given spectrum,
to identify a potential uniquely.
For example, such data can be
a second complete spectrum for different boundary conditions,
knowledge of the potential on a half interval,
or the phase shifts as a function of energy.
However, 
neither a complete spectrum
nor values of potentials or phase shifts for all energies 
can be determined empirically by a finite number of measurements.
Hence, any practical algorithm for reconstructing potentials from data
must rely on additional a priori assumptions, 
if not explicitly then implicitly.
Furthermore, besides energy,
other observables
like particle coordinates or momenta
may have been measured for a quantum system.
Therefore, the approach we study in this Paper is designed
to deal with arbitrary, also non-spectral, data 
and to treat
situation specific a priori information
in a flexible and explicit manner.

Many disciplines contributed
empirical learning algorithms
some of the most widely spread being 
decision trees,
neural networks,
projection pursuit techniques,
various spline methods,
regularization approaches,
graphical models,
support vector machines,
and, becoming especially popular recently,
nonparametric Bayesian methods
\cite{iqs:Wahba:1990,iqs:Hertz-Krogh-Palmer:1991,iqs:Michie-Spiegelhalter-Taylor:1994,iqs:Vapnik:1995,iqs:Lauritzen:1996,iqs:Williams-Rasmussen:1996}.
Motivated by the clear and general framework it provides,
the approach we will rely on
is that of Bayesian statistics
\cite{iqs:Berger:1980,iqs:Robert:1994,iqs:Gelman-Carlin-Stern-Rubin-1995} 
which can easily be adapted to inverse quantum statistics.
Computationally, however, 
its application to quantum systems turns out to be more demanding 
than, for example, typical applications to regression problems.

A Bayesian approach
is based on two probability densities:
1. a likelihood model $p(x|O,v)$,
quantifying the probability of outcome $x$ 
when measuring observable $O$
given a (not directly observable) potential $v$ 
and 
2. a prior density $p_0(v)$ = $p(v|D_0)$
defined over a space ${\cal V}$ of possible potentials
assuming a priori information $D_0$.
Further, let
$D_T$ = $(x_T,O_T)$ = $\{(x_i,O_i)|1\le i\le n\}$
denote available training data
consisting of outcome--observable pairs 
and $D$ = $D_T \cup D_0$
the union of training data and a priori information.
To make predictions
we aim at calculating
the predictive density for given data $D$
\be
p(x|O,D)
= \int \!dv\, p(x|O,v)\, p(v|D).
\ee 
The posterior density $p(v|D)$ appearing in that formula
is related to prior density and likelihood
through Bayes' theorem
$p(v|D)$
=
$p(x_T|O_T,v)\,p_0(v)/p_0(x_T|O_T)$,
where the likelihood
factorizes for independent data
$p(x_T|O_T,v)$ = $\prod_i p(x_i|O_i,v)$
and the denominator is $v$--independent.
The $v$--integral stands for an integral over parameters
if we choose a parameterized space ${\cal V}$,
or for a functional integration
if ${\cal V}$ is an infinite dimensional function space.
Because we will in most cases not be able to solve that
integral exactly we treat it in maximum a posteriori approximation,
i.e., in saddle point approximation if 
the likelihood varies slowly
at the stationary point.
Then, 
assuming
$p(x|O,D) \approx p(x|O,v^*)$
with 
$v^*$ = ${\rm argmax}_{v\in{\cal V}} p(v|D)$
integration is replaced by maximization.

According to the axioms of quantum mechanics
observables $O$ are represented by hermitian operators
and the probability of finding outcome $x$ measuring observable $O$
is given by
\be
p(x|O,v) 
= {\rm Tr} 
\Big(P_O(x) \, \rho(v) \Big)
.
\label{qm-likelihood}
\ee
Here $\rho$ denotes the density operator of the quantum mechanical
system and 
$P_O(x)$ = $\sum_\nu |x_{(\nu)} \!\!><\!\!x_{(\nu)} |$
is the projector into the space spanned 
by the eigenfunctions $|x_{(\nu)}\!\!>$ 
of $O$ having eigenvalue $x$ and which we
assume to be orthonormalized.
If a system is not in an eigenstate of the observable,
a quantum mechanical measurement
changes the state of the system.
Hence, to perform repeated measurements under the same $\rho$
requires the restoration of $\rho$
before each measurement.

In particular, we will consider a canonical ensemble of quantum systems
at temperature $1/\beta$ (setting Boltzmann's constant to 1) 
characterized by the density operator
$\rho$ = 
$\exp (-\beta H)/({\rm Tr\,} \exp(-\beta H))$.
Furthermore, we assume a Hamiltonian 
$H$ = $T + V$ being the sum of a kinetic energy term 
$T$ = $-(1/2m)\Delta$
for a particle with mass $m$
($\Delta$ denoting the Laplacian and $\hbar$ = $1$)
and an unknown local potential
$V(x,x^\prime)$ = $v(x) \delta (x-x^\prime )$
which we want to reconstruct from data.
In case of repeated measurements in a canonical ensemble
one has to wait with the next measurement
until thermal equilibrium is reached again.
We will in the following focus on measurements 
of particle coordinates
in a single particle system in a heath bath with temperature $1/\beta$.
In that case the $x_i$ represent particle positions 
corresponding to measurements of the observable $O_i$ = $\hat x$ 
with $\hat x |x_i\!\!>$ = $x_i|x_i\!\!>$.
The likelihood becomes
\be
p(x_i|\hat x,v) 
=\sum_\alpha p_\alpha |\phi_\alpha(x_i)|^2 
=<|\phi (x_i)|^2>
,
\ee
for
$H|\phi_\alpha\!\!>$ = $E_\alpha |\phi_\alpha\!\!>$
and with $<\!\!\cdots\!\!>$ denoting a thermal expectation 
for probabilities
$p_\alpha$ = 
$\exp(-\beta E_\alpha)/Z$ with
$Z$ = $\sum_\alpha \exp (-\beta E_\alpha)$.
One may also add a classical noise process $p(\bar x_i|x_i)$
on top of $x_i$ in which case an additional integration
is necessary to get the density 
$p(\bar x_i|\hat x,v)$
=
$\int \! dx_i \, p(\bar x_i|x_i) \, p(x_i|\hat x,v)$
of the noisy output $\bar x_i$.

Even at zero temperature
complete knowledge of the true likelihood
would not be enough to determine a potential uniquely.
The situation is still worse in practice 
where only a finite number of probabilistic measurements
is available
and at finite temperatures, 
as the likelihood becomes uniform in the infinite temperature limit.
Hence, in addition to Eq.(\ref{qm-likelihood})
giving the likelihood model of quantum mechanics, 
it is essential to include 
a prior density over a space of possible potentials $v$.
To be able to formulate a priori information explicitly, i.e.,
in terms of the functions values $v(x)$ itself,
we use a stochastic process.
Technically convenient is for example a Gaussian process prior
having the form
\be
p_0(v) 
=
\left(\det \frac{\lambda}{2\pi}{\bf K}_0 \right)^\frac{1}{2}
e^{-\frac{\lambda}{2} 
< v-v_0 | {\bf K}_0 | v-v_0 >}
,
\label{gaussprior}
\ee
with mean $v_0$,
representing a reference potential or template for $v$,
and a
real symmetric, positive (semi--)definite
covariance operator $(\lambda{\bf K}_0)^{-1}$,
acting on potentials $v$ and not on wave functions
and measuring the distance between $v$ and $v_0$.
Hereby $\lambda$ is technically equivalent 
to a Tikhonov regularization parameter.
Typical choices for ${\bf K}_0$ implementing smoothness priors
are the negative Laplacian
${\bf K}_0$ = $-\Delta$
or a Radial Basis Function prior
${\bf K}_0$ = $\exp{(-{\sigma_{\rm RBF}^2}{\Delta}/2)}$
\cite{iqs:Girosi-Jones-Poggio:1995}.
More general, Gaussian process priors 
can be related to general approximate symmetries.
Assume, for example,
we expect the potential to commute approximately with a 
unitary symmetry operation $S$. Then,
$V \approx S^\dagger V S$ = ${\bf S} V$,
defining an operator ${\bf S}$ acting on potentials $V$.
In that case a natural prior would be
$p_0\propto\exp (-E_{S})$ with
$E_S$ 
= $<\!V-{\bf S}V|V-{\bf S}V\!>\!/2$
= $<\! V |{\bf K}_{0}|V\!>\!/2$
for
${\bf K_0}$ = 
$({\bf I}-{\bf S})^\dagger ({\bf I}-{\bf S})$
and ${\bf I}$ denoting the identity.
Note that symmetric potentials are in the null space of 
such a ${\bf K}_0$
so also another prior has to be included
if not already the combination with training data does
determine the potential.
Similarly, for a Lie group 
${\bf S}(\theta)$ = $\exp (\theta {\bf s})$
an approximate infinitesimal symmetry is implemented by
${\bf K_0}$ = ${\bf s}^\dagger{\bf s}$.
In particular, a negative Laplacian smoothness prior
enforces approximate symmetry under infinitesimal translation.
Alternatively, 
a more explicit prior implementing an approximate
symmetry can be obtained by choosing a symmetric reference potential
$V_S$ = $S^\dagger V_S S$
and $E_S$ = $<\!V-V_S|V-V_S\!>\!/2$.

While a Gaussian process prior is only able to
model a simple concave prior density,
more general prior densities can be 
arbitrarily well approximated by mixtures
of Gaussian process priors \cite{iqs:Lemm:1999}
\be
p_0(v) 
=\sum_k^M p_k \,p_0(v|k),
\label{gauss-mixture}
\ee
with mixture probabilities $p_k$
and Gaussian prior processes $p_0(v|k)$
having means $v_k$ and covariances $(\lambda {\bf K}_k)^{-1}$,
where $\lambda$ plays the role of a ``mixture temperature''.
Similarly,
one may introduce so called hyperparameters
which parameterize the prior density \cite{iqs:Lemm:1999}.
Those hyperparameters have to be included in the set
of integration variables to obtain the predictive density
and can also be treated in maximum a posteriori approximation.

To find the potential with maximal posterior
we have to maximize 
\be
p(v|D) 
\propto 
p_0(v) \prod_i^n p(x_i|O_i,v),
\label{posterior}
\ee
where we assumed independent training data.
This can be done by setting the functional derivatives
of the log--posterior
(technically often more convenient than the posterior) 
with respect to $v(x)$ to zero, i.e., 
\be
\delta_{v(x)} \ln p(v|D) 
= 0
\label{stat-eq}
\ee
where  $\delta_{v(x)}$ stands for $\delta/\left(\delta v(x)\right)$.
To calculate the functional derivative 
of the likelihoods in Eq.(\ref{posterior})
we require
$\delta_{v(x)} E_\alpha$ 
as well as 
$\delta_{v(x)} \phi_\alpha(x^{\prime})$.
Those can be found by taking the functional derivative
of the eigenvalue equation of the Hamiltonian yielding,
for orthonormal eigenfunctions, 
\bea
\delta_{v(x)} E_\alpha 
&=& <\!\!\phi_\alpha |\, \delta_{v(x)} H \,| \phi_\alpha\!\!>
=|\phi_\alpha(x)|^2
,
\label{deltaE-nonp}
\\
\delta_{v(x)} \phi_\alpha(x^{\prime})
&=& \sum_{\gamma\ne \alpha} \frac{1}{E_\alpha-E_\gamma} 
\phi_\gamma(x^{\prime})\phi^*_\gamma(x) \phi_\alpha (x)
,
\eea
using the fact that 
$\delta_{v(x)} H (x^\prime,x^{\prime\prime})$
= $\delta_{v(x)} V (x^\prime,x^{\prime\prime})$
= $\delta(x-x^\prime) \delta (x^\prime-x^{\prime\prime})$.
Now it is straightforward to calculate the functional derivative 
of the likelihood 
\bea
&&
\delta_{v(x)}p(x_i|\hat x,v)
=
\\&&
\quad
<\left(\delta_{v(x)}\phi^*(x_i)\right) \phi (x_i)> 
+<\phi^*(x_i)\delta_{v(x)}\phi (x_i)> 
\nonumber\\&&
-
\beta \left(
< |\phi (x_i)|^2 |\phi (x)|^2 >
-< |\phi (x_i)|^2>< |\phi^2 (x)|^2>
\right)
.
\nonumber
\eea
Finally, using
\be
\delta_{v(x)} \ln p_0 
= -\lambda{\bf K}_0(v-v_0)
,
\label{prior-dev}
\ee
the functional derivative of the posterior can be calculated.
The formula (\ref{prior-dev}) is also valid for 
Gaussian mixture models of the form (\ref{gauss-mixture})
provided we understand
${\bf K}_0$ 
=
$\sum_k p_0(k|v)\,{\bf K}_k$
and $v_0$ 
= 
${\bf K}_0^{-1} \sum_k p_0(k|v)\, {\bf K}_k v_k$
where $p_0(k|v)$ = ${p_0(v|k)p_0(k)}/{p_0(v)}$.

It is straightforward to include
also other kinds of data or a priori information.
For example, 
a Gaussian smoothness prior as in Eq.(\ref{gaussprior})
with zero reference potential $v_0\equiv 0$
and, say, $v(x)$ = 0 at the boundaries tends to lead 
to flat potentials when the regularization parameters $\lambda$ becomes large.
For such cases it is useful to include besides smoothness
also a priori information or data
which are related more directly to the depth of the potential.
One such possibility is to include information 
about the average energy 
$U$ 
= $<\!E\!>$
=$\sum_\alpha p_\alpha E_\alpha$.
The average energy can then be controled by introducing a Lagrange multiplier
term $\mu (U - \kappa)$,
or, technically sometimes easier,
by a term representing noisy energy data 
\be
p_U \propto e^{-E_U}
,\quad
E_U =
\frac{\mu}{2} (U - \kappa)^2,
\label{averageE-penal}
\ee
so that $U\rightarrow\kappa$ for
$\mu\rightarrow\infty$.
Using 
$\delta_{v(x)} U$ =
$<\!\delta_{v(x)} E\!>-\beta <\!E\; \delta_{v(x)} E\!>$
+ $\beta <\!E\!> <\!\delta_{v(x)} E\!>$,
we find for the functional derivative of $E_U$
\bea
\delta_{v(x)} E_U 
&=&
  \mu\left(U-\kappa\right)
  \Big(<\!|\phi (x)|^2\!>
\\&&
  -\beta \left( <\!E \; |\phi (x)|^2\!>
  -U <\!|\phi (x)|^2\!> 
\right)
\Big)
.
\nonumber
\eea

Collecting all terms we are now able to solve
the stationarity Eq.(\ref{stat-eq}) by iteration.
Starting with an initial guess $v^{(0)}$,
choosing a step width $\eta$
and a positive definite iteration matrix ${\bf A}$
we can iterate according to
\bea
v^{(k+1)}
 &=&
v^{(k)}\! +
\eta {\bf A}^{-1}
\Big(
{\bf K}_0^{(k)} (v_0\!-\!v^{(k)})
\label{iter1}
\\&&
+\sum_i \delta_x\ln p(x_i|\hat x,v^{(k)})
-\delta_x E_U^{(k)}
\Big)
.
\nonumber
\eea
Here we included an 
$E_U$ term which depends on $v$,
like ${\bf K }_0$ for mixture models,
and thus changes during iteration. 
Typically, $\eta$ and often also ${\bf A}$ are
adapted during iteration.
In the numerical examples we have studied
${\bf A}$ = $\lambda {\bf K}_0$
proved to be a good choice.

The numerical difficulties of 
the nonparametric Bayesian approach 
arises from the fact that 
the quantum mechanical likelihood (\ref{qm-likelihood})
is non--Gaussian and non--local in the potential $v(x)$.
Thus, even for Gaussian priors 
no $v(x)$--integration can be carried out analytically.
In contrast, for example,
Gaussian regression problems
have a likelihood being Gaussian and local in the function of interest,
and an analogous nonparametric Bayesian approach 
with a Gaussian process prior
requires only to deal with matrices with dimension not larger than
the number of training data \cite{iqs:Williams-Rasmussen:1996}.
The following examples will show, however, 
that a direct numerical solution of Eq.(\ref{stat-eq})
by discretization is feasible for one--dimensional problems.
Higher dimensional problems, on the other hand,  
require further approximations.
For example, treating inverse many--body problems 
on the basis of a Hartree--Fock approximation
is ongoing work.

\begin{figure}
\vspace{0.2cm}
\begin{center}
\epsfig{file=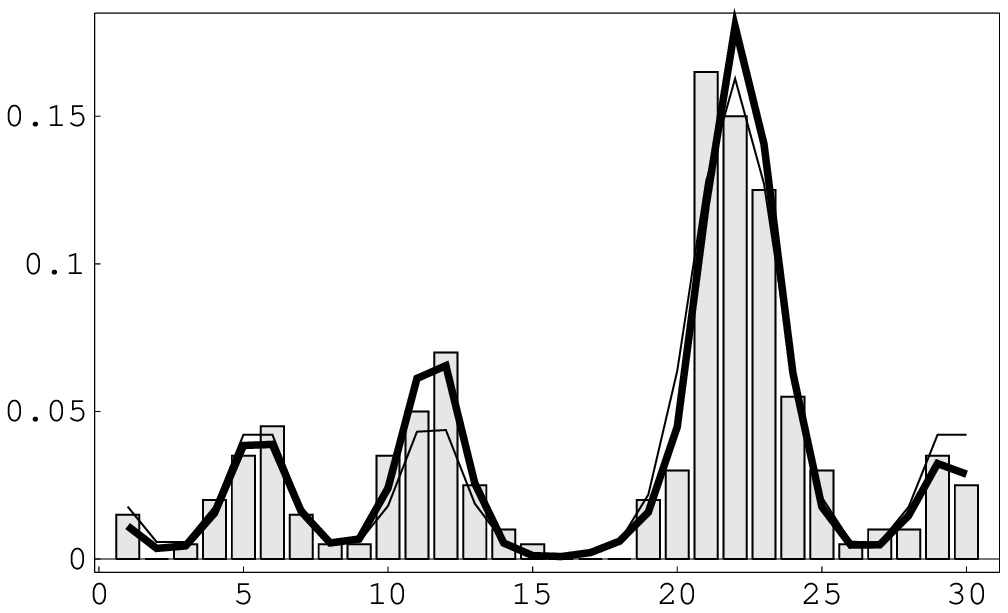, width= 40mm}
$\quad$
\epsfig{file=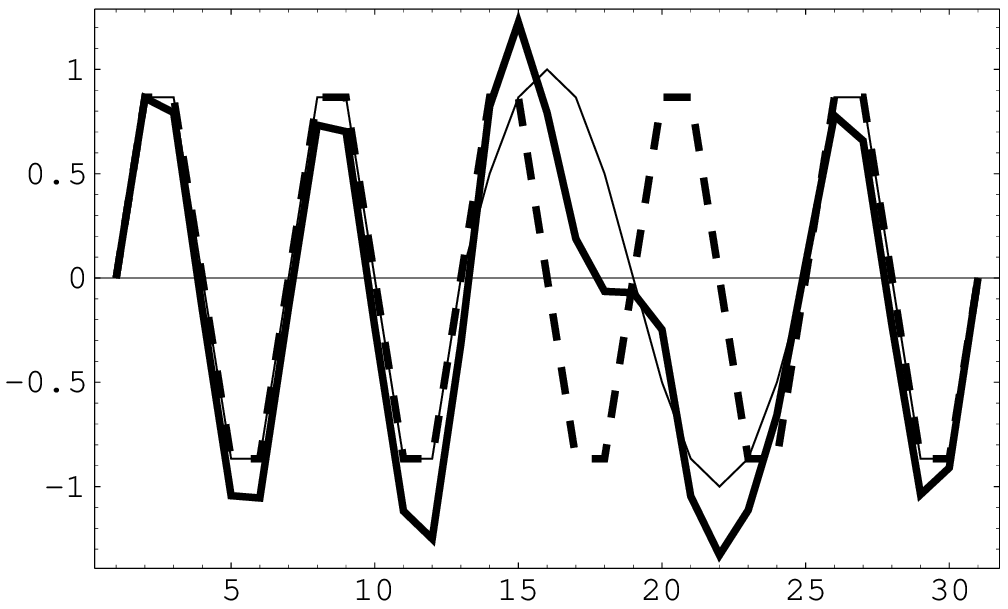, width= 40mm}
\setlength{\unitlength}{1mm}
\begin{picture}(0,0)
\put(-21,31){\makebox(0,0){(a) likelihoods}}
\put(25,31){\makebox(0,0){(b) potentials}}
\end{picture}
\vspace{-0.4cm}
\end{center}
\caption{Reconstruction of an approximately periodic potential
from coordinate measurements.
(a) shows 
empirical density (bars),
true likelihood (thin line),
and reconstructed likelihood (thick line).
(b) shows the true potential (thin line),
reference potential $v_0$ = $\sin(\pi x/3)$ (dashed line),
and reconstructed potential $v$ (thick line).
(200 data points, mesh with 30 points,
$m$ = 0.25, 
$\beta$ = 4,
${\bf K}_0$ = $-\Delta$,
$\lambda$ = 0.2,
$\mu$ = 0,
$U (v_{\rm true})$ = $-0.354$
resulting in
$U(v)$ = $-0.552$.)
}
\label{periodic-fig}
\end{figure}

\begin{figure}
\begin{center}
\epsfig{file=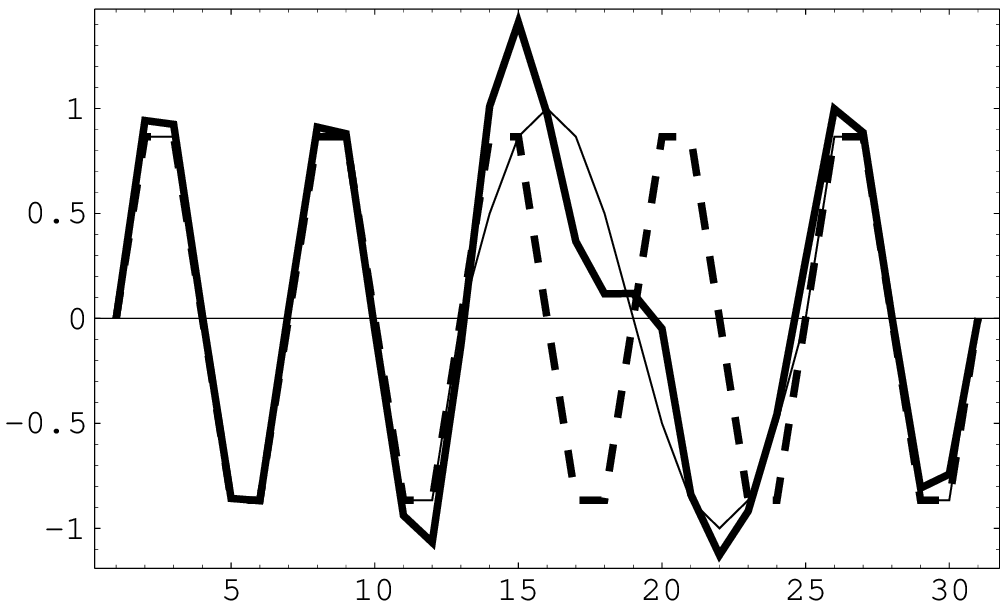, width= 40mm}
$\quad$
\epsfig{file=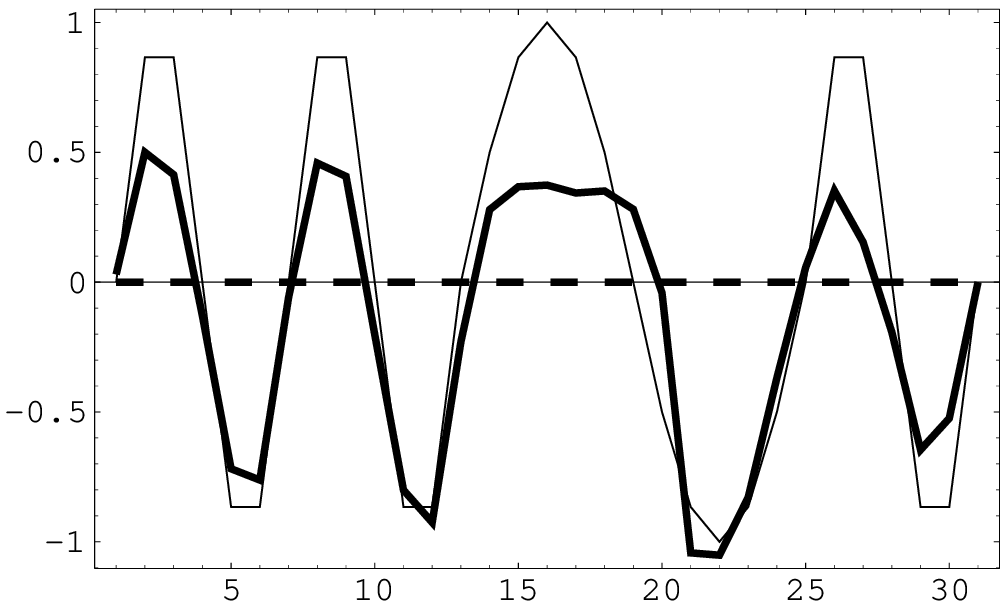, width= 40mm}
\setlength{\unitlength}{1mm}
\begin{picture}(0,0)
\put(-21,31){\makebox(0,0){(a) potentials}}
\put(25,31){\makebox(0,0){(b) potentials}}
\end{picture}
\vspace{-0.4cm}
\end{center}
\caption{Reconstruction with energy penalty term $E_U$
with $\kappa$ = $U (v_{\rm true})$ = $-0.354$.
Shown are true potential (thin lines),
reconstructed potential (thick lines),
reference potential (dashed).
(a) 
$K_0$ = $-\Delta$, 
$v_0(x)$ = $\sin(\pi x/3)$, 
$\mu$ = 1000, 
$U(v)$ = $-0.353$.
(b) 
${\bf K}_0$ = $-\Delta -\gamma \Delta_T$,
$v_0\equiv 0$, 
$T$ = 6,
$\gamma$ = 0.12,     
$\lambda$ = 0.05,  
$\mu$ = 10, 
$U(v)$ = $-0.351$.
All other parameters as for Fig.\ref{periodic-fig}.
}
\label{periodic-fig2}
\end{figure}

\begin{figure}
\begin{center}
\epsfig{file=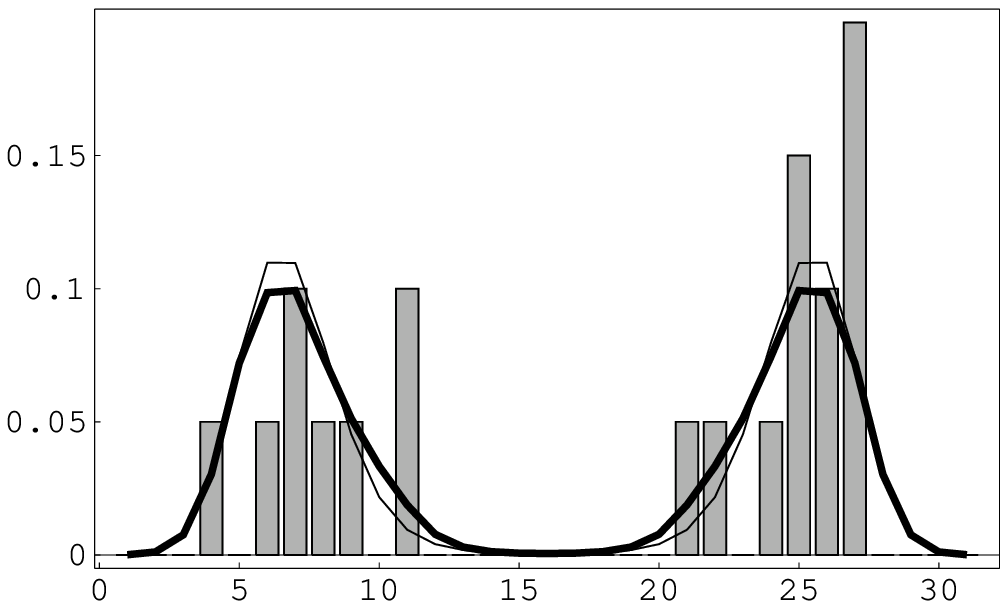, width= 40mm}
$\quad$
\epsfig{file=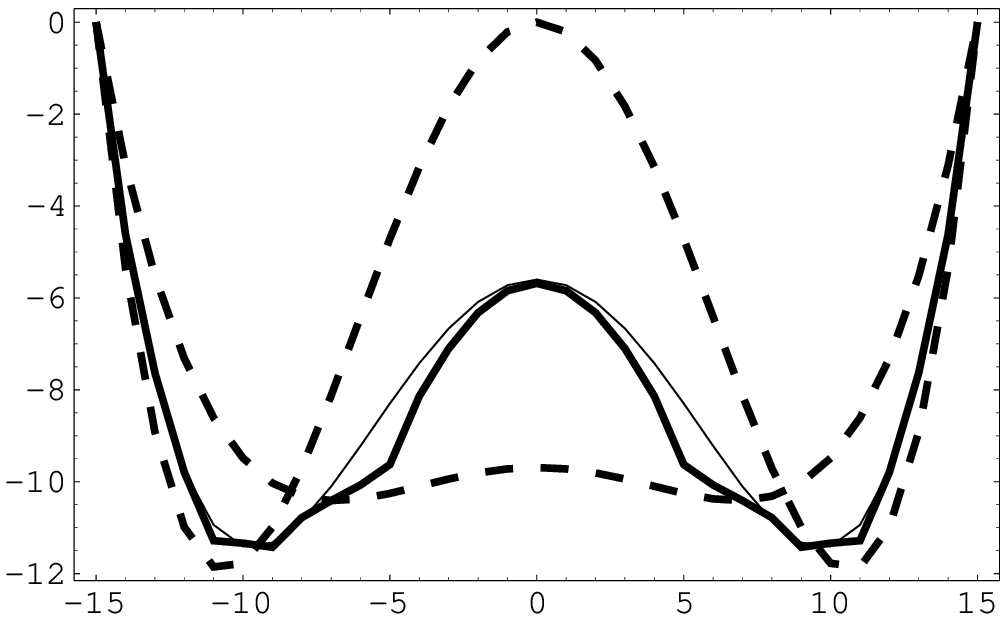, width= 40mm}
\setlength{\unitlength}{1mm}
\begin{picture}(0,0)
\put(-21,31){\makebox(0,0){(a) likelihoods}}
\put(25,31){\makebox(0,0){(b) potentials}}
\end{picture}
\vspace{-0.4cm}
\end{center}
\caption{Reconstruction of symmetric potential
with mixture of Gaussian priors.
(a) shows 
empirical density (bars),
true likelihood (thin line),
and reconstructed likelihood (thick line).
(b) shows the true potential (thick line),
the two reference potential $v_1$ and $v_2$ (dashed lines),
and the reconstructed potential $v$ (thick line).
(20 data points, mesh with 31 points,
for $v(x)$ = $v(-x)$ and 
boundary condition $v(\pm 15)$ = 0,
with 
$m$ = 0.1,
$\beta$ = 1, 
${\bf K}_0$ = $-\Delta$,
$\lambda$ = 0.5,
$\mu$ = 10
$\kappa$ = $-9.66$ = $U(v_{\rm true})$ resulting in
$U(v)$ = $-10.1$.
)
Because both reference potentials are partially 
supported by the data 
the approximated $v$ 
is essentially
a smoothed mixture between $v_1$ and $v_2$
in regions where no data are available
with mixture coefficients for prior components 
$p_0(1|v)$ = 0.57, $p_0(2|v)$ = 0.43.
}
\label{pot-fig}
\end{figure}

In the following numerical examples we discuss the reconstruction 
of an approximately periodic, one--dimensional potential.
As a first example, assume we expect $v$
to be similar to a periodic reference potential $v_0$,
which we choose as reference potential.
To enforce the deviation from $v_0$ to be smooth
we take as prior on $v$ a negative Laplacian covariance, i.e.,
${\bf K}_0$ = $-\Delta$.
Fig.\ref{periodic-fig} shows representative numerical results
for a grid with 30 points and
200 data sampled
from a true likelihood $p(x|\hat x,v_{\rm true})$
corresponding to a true potential $v_{\rm true}$.
The reconstructed potential $v$ have been found
by iterating without energy penalty term $E_U$
according to Eq.(\ref{iter1}) with
${\bf A}$ = ${\bf K}_0$.
Choosing zero boundary conditions for $v$ the matrix ${\bf K}_0$ is invertible.
Consistent with the boundary conditions on $v$
we took periodic boundary conditions for the eigenfunctions $\phi_\alpha$.
Note that the data have been sufficient to identify clearly
the deviation from the periodic reference potential.
Fig.\ref{periodic-fig2}a
shows the same example
with an energy penalty term $E_U$ with $\kappa$ = $U(v_{\rm true})$.
While the reconstructed likelihood (not shown) is not much altered,
the true potential is now better approximated in regions where it is small.

Fig.\ref{periodic-fig2}b
shows the implementation of approximate periodicity
by an operator $\Delta_T$
measuring the difference between the potential
$v(x)$ and the potential translated by $T$ and 
and which is defined by 
$<\!\!v|\Delta_T |v\!\!>$ = $\int \!dx |v(x)-v(x+T)|^2$
for periodic boundary conditions on $v$.
To find smooth solutions
we added a negative Laplacian term with zero reference potential, i.e.,
we used
${\bf K}_0$ = $-\Delta -\gamma \Delta_T$.
To have an invertible matrix for periodic boundary conditions on $v$
we iterated this time with
${\bf A}$ = ${\bf K}_0$ + $0.1 {\bf I}$.
The implementation of approximate periodicity by
$\Delta_T$ instead of an periodic $v_0$
is more general in as far it allows arbitrary functions
with period $T$.
As, however, 
the reference function of the Laplacian term does not fit the true
potential very well, the reconstruction is poorer
in regions where the potential is large
and thus no data are available.
In these regions a priori information is of special importance.
Finally, Fig.\ref{pot-fig}
shows the implementation of a mixture of Gaussian prior processes.

In conclusion, 
we have applied 
a nonparametric Bayesian approach to inverse quantum statistics
and shown its numerical feasibility for one--dimensional examples.

\bibliographystyle{prsty}
\bibliography{iqslanl}

\end{document}